\documentstyle[twoside,fleqn,espcrc2,epsfig,psbox]{article}

\newcommand{\AmS}{{\protect\the\textfont2
  A\kern-.1667em\lower.5ex\hbox{M}\kern-.125emS}}

\title{Gamma-Ray Burst Spectra and Time Histories From  2 to 400 keV}

\author{E. E. Fenimore\\MS D436, Los Alamos National Laboratory,
        Los Alamos, NM 87544}
       
\begin{document}

\begin{abstract}
The Gamma-Ray burst detector on {\it Ginga} consisted of a proportional
counter to observe the x-rays and a scintillation counter to observe
the gamma-rays. It was ideally suited to study the x-rays associated
with gamma-ray bursts (GRBs).  {\it Ginga} detected $\sim120$ GRBs and
22 of them had sufficient statistics to determine spectra from 2 to
400 keV.  Although the {\it Ginga} and BATSE trigger criteria were
very similar, the distribution of spectral parameters was different.  
{\it Ginga} observed bend energies in the spectra
down to 2 keV and had a larger fraction of bursts with low energy
power law indexes greater than zero. The average ratio of energy in
the x-ray band (2 to 10 keV) compared to the gamma-ray band (50 to 300
keV) was 24\%.  Some events had more energy in the x-ray band than in
the gamma-ray band.  One {\it Ginga} event had a period of time
preceding the gamma rays that was effectively pure x-ray emission.
This x-ray ``preactivity'' might be due to the penchant for the GRB time
structure to be broader at lower energy rather than a different
physical process.  The x-rays tend to rise {\it and} fall slower than
the gamma rays but they both tend to peak at about the same time.
This argues against models involving the injection of relativistic
electrons that cool by synchrotron radiation.
\end{abstract}


\maketitle

\def\M{{\cal M}}                      %  stellar mass
\def\Mbol{M_{\rm bol}}              %  abs. bolometric magn.
\def\sun{_\odot}                    %  sub-sun symbol
\def\msun{{\cal M}_\odot}                  %  M(sun)
\def\mmsun{{\rm M/M}_\odot}               %  M/M(sun)
\def\rsun{{\rm R}_\odot}                  %  R(sun)
\def\rrsun{{\rm R/R}_\odot}               %  R(sun)
\def\Teff{{\it T}\lower.5ex\hbox{\rm eff}}
\def\degrees{$^\circ$}
\def\arcmin{$^\prime$}
\def\arcsec{$^{\prime\prime}$}
\def\hrs{\raise.5ex\hbox{h}}
\def\min{\raise.5ex\hbox{m}}
\def\sec{\raise.5ex\hbox{s}}
%
%  Mathmatical definitions
%
\def\twiddle{$\sim$}                % for use in open text
\def\mean#1{\langle #1 \rangle}     % this and the next two need 
\def\lsim{<\kern-1.7ex $\lower0.85ex\hbox{$\sim$}$\ }     % math surrounds.
\def\gsim{>\kern-1.7ex $\lower0.85ex\hbox{$\sim$}$\ }     % math surrounds.
\def\element#1 #2{${}^{#1}$#2}
%
%  Personal definitions
%  
\def\MH{{M\lower.5ex\hbox{H}}}
\def\MHe{{M\lower.5ex\hbox{He}}}
\def\Mstar{{M\lower.5ex\hbox{$\star$}}}
\def\etal{{\it et al\/}.}
\def\ie{{\it i.e\/}.}
\def\eg{{\it e.g\/}.}    
\def\vs{{\it vs\/}.}
\def\cf{{\it cf}.}
\def\gmode{\hbox{{\it g}-mode}}
\def\Gyr{{\rm Gyr}}
\def\lhat{l(l+1)}
\def\etc{{\it etc.}}
\def\del{\nabla}
\section{INTRODUCTION}

Gamma-ray bursts (GRBs) are aptly name, most of their power is usually
emitted in the 100 keV to 1 MeV energy range.
Early results indicated that only a few percent of the energy of a Gamma-Ray
burst (GRB) occurs in the 2-to-10 keV x-ray range \cite{laros84,katoh84},
although  the X-ray emission might outlast the main 
GRB event in some bursts (so called X-ray tails).
Based on 
these intriguing  results, the gamma-ray burst 
detector (GBD) flown aboard the {\it Ginga} satellite 
was specifically designed 
to investigate burst spectra in the X-ray regime \cite{mur89}. 
{\it Ginga} was launched in 
February of 1987, and the GBD was operational from March, 1987, until the 
reentry of the spacecraft in October, 1991.

More recently, the BeppoSax satellite has observed several bursts in both
x-rays and gamma-rays \cite{piro97,piro98,frontera98} and 
has discovered soft x-ray
afterglows \cite{costa97}.  
This latter discovery has opened the way for the
long-sought GRB counterparts, including one with a measured redshift
\cite{metz97}.

Here, we summarize x-ray results from the {\it Ginga} experiment.

\section{INSTRUMENTAL DETAILS}
The GBD aboard {\it Ginga} consisted of a proportional 
counter (PC) sensitive to
photons in the 2-to-25 keV range and a scintillation counter (SC) recording 
photons with energies between 15-to-400 keV. Each detector had an $\approx60$ 
cm$^2$ effective area. The PC and SC provided 16 and 32 channel spectra, 
respectively, over their indicated energy ranges. The detectors were 
uncollimated except for the presence of shielding to reduce backside 
illumination and the mechanical support for the PC window.
The field of view was effectively  $\pi$ steradian for both
the PC and the SC.

Each detector produced both spectral and time history-data for GRBs.
Photons within the  energy range of each detector were
accumulated into time samples. The hardware ensured that the
 time samples for both 
detectors started and stopped at the same time. This facilitates measuring the
relative timing between the x-rays and gamma-rays.
The temporal resolution of the time-history data 
depended on the telemetry mode.  An on-board trigger system checked the 
time histories for a significant increase ($11\sigma$ on either 1/8, 1/4, or 
1 s time scales).  Upon such a trigger, special high resolution temporal
data (called ``Memory Read Out'' [MRO]) would be produced.  The MRO data
has time resolution of 0.03125 s.  The PC MRO time history
extends from 32 s before the
trigger to 96 s after the trigger.  The SC MRO time history extends
from 16 s before the trigger to 48 s after the trigger.  During the
period when both detectors have MRO time histories, the MRO samples
were in phase with each other.
In addition to the MRO data, ``Real Time'' data were often 
continuously available.
Three different telemetry modes resulted in PC and SC time histories with
either 0.125, 1.0, or 4.0 s time resolution.

In
burst mode, the GBD recorded spectral data from the 
PC and SC at 0.5 s intervals 
for 16 s before the burst trigger time and for 48 s after the
trigger. The PC and SC had 16 and 32 energy channels, respectively.
The MRO data were used for 
many of the spectral fits 
described here. In the event that MRO data were not available for a burst, we
utilized the spectral data from the ``real time'' telemetry modes.
 For these bursts, spectral data were
available with either 2, 16, or 64 s accumulations.  For the longer
accumulations, spectral 
studies were not generally feasible.
 See \cite{mur89} and \cite{stroh98}
for more information concerning the instrument.

The {\it Ginga} GBD was in operation from March 1987 to October 1991. During 
this time $\approx120$ $\gamma$-ray bursts were identified 
\cite{oga91,fen93}. 
{}From this group,  a sample of 22 events with
statistically significant spectral data were analyzed \cite{stroh98}.
These events occurred within the 
forward, $\pi$ steradian field of view of the detectors (front-side events). 
Front-side events are easily identified because they have consistent fluxes
 in the
energy range observed by both instruments (15 to 25 keV).
Excluded bursts usually show strong rolloffs below 25 keV
and spectral fits with an absorption component due to aluminum 
(a principal spacecraft material) are consistent with this interpretation.

Because the incidence angle was known for just four of the events, we
had to assume an incidence angle for the remaining events. 
We selected 37 degrees for the incidence angle when the angle was unknown.
This is a typical angle considering that the mechanical support
for the window
on the PC acts as
a collimator limiting the field of view to
an opening angle of
$\sim\pm 60$ degrees.
Extensive simulations \cite{stroh98} demonstrated that our conclusions
were not affected by the uncertainty in the incidence angle.

We have adopted the spectral 
model employed by \cite{band93} because of its 
relative simplicity and ability to 
accurately characterize a wide range of spectral continua. In addition, this 
choice facilitates direct comparison of our results with those 
obtained from
bursts seen by the Burst and Transient Source Experiment (BATSE).
This model has the form
$$
N(E) = A E^{\alpha} \exp (-E/E_0)
$$
if
$$
(\alpha -\beta)E_0 \ge E
$$
and
$$
N(E) = A\left [ (\alpha -\beta)E_0 \right ] ^{\alpha
-\beta} \exp (\beta -\alpha) E^{\beta}
$$
if
$$
(\alpha -\beta)E_0 \le E .
$$
Here, $A$ is an overall scale factor, $\alpha$ is the
low-energy spectral index,
$\beta$ is the
high-energy spectral index, and $E_0$ is the
exponential cutoff
or bend energy.
The model parameters 
are adjusted iteratively until a minimum in $\chi^2$ is obtained.

\section{X-RAY CHARACTERISTICS OF\\{\it {\bf GINGA}} BURSTS}

Of particular interest is the behavior of the {\it Ginga} 
sample at x-ray energies. About 40\%  of the bursts in the sample show a 
positive spectral number index below 20 keV (i.e., $\alpha >
0$),
 with the suggestion of 
 rolloff toward lower energies in a few of the bursts ($\alpha$ as
large as $+1.5$) \cite{stroh98}.
Unfortunately, the 
lack of data below 1 keV, and the often weak signal 
below 5-to-10 keV precludes  
 establishing the physical process (photoelectric absorption,
self-absorption) that may be 
involved in specific bursts. 
Observations of the low-energy asymptote can place serious constraints
on several GRB models, most notably the synchrotron shock model which
predicts that $\alpha$ should be between -3/2 and -1/2 \cite{katz94}.
Crider et al.
\cite{crider97}  uses examples from BATSE to argue that some GRB
violate these limits during some {\it time-resolved} samples.  We find
violations of these limits in the {\it time-integrated} events.

In our sample of bursts, we find that $\alpha$ can be both less than
zero and greater than zero.  Negative $\alpha$'s are often seen in
time-integrated BATSE spectra.  Positive $\alpha$'s for which the spectrum
rolls over at low energies are usually only seen in {\it
time-resolved} BATSE spectra \cite{crider97}.

Our distribution of parameters ($\alpha, \beta, E_0$) is broader than
that found by BATSE.  For example, 40\% of our events had $\alpha$
greater than 0 compared to  only 15\% of BATSE events, and we had bend
energies that occurred as low as 1.7 keV, wheras the lowest BATSE
bend energy was 15 keV.
The {\it Ginga} trigger range (50 to 400 keV) was virtually the same
as BATSE's. Thus, we do not think we are sampling a different population of
bursts, yet we get a different range of fit parameters.
One possible explanation might be that GRBs have two break energies, one
often in the 50 to 500 keV range and the other near 5 keV.  Both BATSE and
{\it Ginga} fit with only a single break energy so BATSE tends to find breaks
near the center of its energy range, and we tend to find
breaks in our energy range.
Without good high energy observations of bursts with low $E_0$, it is
difficult to know whether they also have a high-energy bend.

\section{X-RAY EMISSION RELATIVE TO GAMMA-RAY EMISSION}

Early measurements of the x-rays associated with gamma-rays were 
fortuitous observations by collimated x-ray detectors that just happen
to catch a GRB in their field of view.  From a few events it appeared
that the amount of energy in x-rays was only a few percent (Laros et
al., 1984 \cite{laros84}) 
confirming that GRBs were, indeed, a gamma-ray phenomena. 
  The {\it Ginga} experiment was designed with a wide field
of view to detect a sufficient number of events to determine the range
of x-ray characteristics.  Early reports from {\it Ginga} events
indicated that sometimes a much larger fraction of the emitted energy
was contained in the x-rays. For example, by comparing the signal in
the proportional counter (roughly 2 to 25 keV) to that of the
scintillator (roughly 15 to 400 keV), we reported an x-ray to
gamma-ray emission ratio up to $\sim $46\% (Yoshida et al., 1989 \cite{yos89}).
  Such a ratio depends on the
bandpass for which it is evaluated.  
Using the best fit $\alpha, \beta,$ and $E_0$ from \cite{stroh98}, one can
find the ratio of emission for a typical x-ray
bandpass (2 to 10 keV) compared to the BATSE energy range (50 to 300
keV).
The ratio is defined to be
$R_{x/g} = \int_{2}^{10}EN(E)dE/\int_{50}^{300}EN(E)dE$.
  Figure 1
presents the distribution for the 22 events analyzed in \cite{stroh98}.
Although the ratio is often a few percent, for some events the ratio
is near (or larger than) unity.
 Some GRBs actually have more energy
in the x-ray bandpass than the gamma-ray bandpass. 
The simple average of the 22 values is 24\%.
This large value arises because of the few events with nearly equal energy
in the x-ray and gamma-ray bandpass.  However, even the logarithmic average
is 7\%.

\begin{figure}[tb]
%\centerline{\epsfig{file=ginga_rev_figure7.eps,height=2.100in,clip=}}
\centering
\psbox[xsize=0.35#1,ysize=0.35#1]{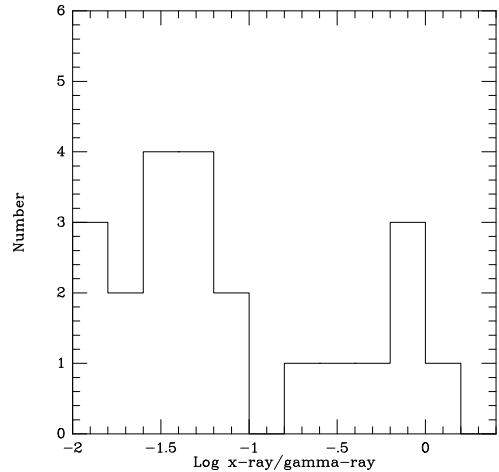}
\caption{
The distribution of the ratio of the energy emitted in
x-rays relative to that emitted in gamma-rays.  The x-ray bandpass is defined
to be from 2 to 10 keV, and the gamma-ray bandpass is the BATSE range of
50 to 300 keV.  Note there are some examples of equal energy in the x-rays
and the gamma-rays.
The average ratio of the energy in the x-rays to the energy in the gamma-rays
is 24\%. From Strohmayer et al., 1998.}
\label{fig:fig1}
\end{figure}

\section{X-RAY PREACTIVITY OR PULSE SPREADING?}

It is common for the x-ray emission to last longer than the gamma-ray
emission \cite{piro97,piro98,frontera98,yos89,mur91,mur92}.
There is one clear example of x-ray activity preceding the GRB \cite{mur91}.
Figure 2 presents the time history of GB900126 in several energy
bands.  From top to bottom, one sees an x-ray hardness ratio, the count
rate in the PC (1-to-28 keV, 0.03125 s resolution), and five time histories
based on the PC and SC  pulse height analysis (PHA) data (0.5 s resolution).
The horizontal arrow in the bottom panel indicates the period that is 
effectively pure x-rays.  During the period marked by the arrow, there is
no detectable emission above 7 keV, but there is significant emission below
7 keV (see Murakami et al. \cite{mur91}).

\begin{figure}[tb]
\centerline{\epsfig{file=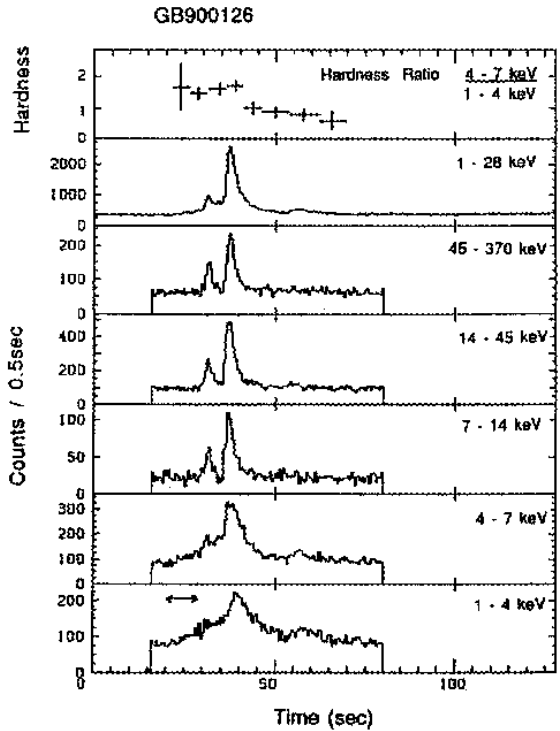,height=3.875in,clip=}}
\caption{The temporal evolution of GB900126 demonstrating the x-ray
``preactivity.''  The top panel is the x-ray hardness and the next
panel is the total count rate in the proportional counter.  The next 5
panels use the energy-resolved PHA data in 5 energy bands from 1 to
370 keV.  Note that the peak widths become wider with lower energy but
they are not
substantially shifted in time. Synchrotron cooling of ejected
electrons is expected to produce substantial shifts at lower energy.
In the bottom panel, the time period mark with a horizontal arrow is
the period of x-ray preactivity during which the emission occurs only at
energies less than 7 keV. (From Murakami et al., 1989.)
}
\label{fig:fig2}
\end{figure}

It would be misleading to refer to the x-ray phase as a ``precursor'' 
because it is not a separate peak.  ``Preactivity'' gives the connotation
of a separate cause for the emission although not necessarily as 
distinct as ``precursor'' might imply. However, the x-ray phase in Figure 2
could be due to the penchant for peaks to be wider at low energy.
Figure 3 shows the average autocorrelation of GRB time histories in
four energy bands.  At higher energy, the autocorrelation is
wider. This spreading of the time structure with energy is also seen
in the average rise and fall times of individual pulses \cite{fen95}.
On average, the peak temporal width  scales as $E^{-0.45}$.  The
spreading in Figure 2 is larger but that could reflect that some 
bursts have larger spreading than the average.

Perhaps there is not a separate cause for the x-ray emission, but rather, the 
physics responsible for the spreading as a function of energy causes the 
x-rays to appear to turn on before the gamma-rays.

The physics responsible for the spreading is unclear.  Kazanas,
Titarchuk, \& Hua (1998) \cite{kazanas98} have 
suggested two processes, synchrotron
cooling of injected relativistic electrons and Compton downscattering of
injected photons.  Synchrotron cooling of
ejected electrons can produce a time structure that scales as
$E^{-1/2}$. However, such cooling should cause the emission at lower energy
to peak later.  In most GRBs, the rise and the fall are slower at low
energy (causing the $E^{-0.45}$ spreading) but the peak of the
emission is not delayed substantially.  In the calculations of Kazanas,
Titarchuk, \& Hua, the low energy emission (e.q., 25-to-50 keV) starts
to rise {\it after} the high energy emission (e.q., 300-t0-1000 keV)
has completely fallen.  Because GRBs never seem to do this, we 
think it strongly argues against
synchrotron cooling as the mechanism that produces the peaks in GRBs.

\begin{figure}[tb]
%\centerline{\epsfig{file=rome_autocorr_fig.eps,height=3.875in,clip=}}
\centering
\psbox[xsize=0.4#1,ysize=0.4#1]{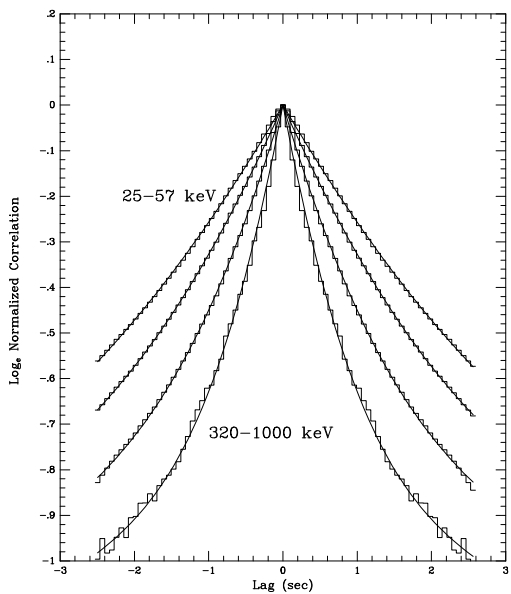}
\caption{Average autocorrelation of 45 bright BATSE gamma-ray bursts
in four energy channels.  At higher energy, GRBs have shorter
timescales.  The solid curves are fits of the sum of two exponentials
to the autocorrelation histogram. (From Fenimore et al. 1995.)}
\label{fig:fig3}
\end{figure}

\end{document}